\documentclass[conference]{IEEEtran}

\usepackage{amsthm}
\usepackage{amsmath}

\usepackage{algorithm,algpseudocode}
\usepackage[dvips]{graphicx}
\usepackage{amssymb}
\usepackage{multirow}
\usepackage{colortbl}
\usepackage{array, tabularx}
\usepackage{subfigure}
\usepackage{color}
\usepackage{graphicx}

\ifCLASSINFOpdf
\else
\fi
\begin{document}
%
\title{When Pilots Should Not Be Reused Across Interfering Cells in Massive MIMO}

\author{\IEEEauthorblockN{Ji Yong Sohn, \textit{Student Member, IEEE}, Sung Whan Yoon, \textit{Student Member, IEEE,}\\ 
and Jaekyun Moon, \textit{Fellow, IEEE}}
\\[1pt]
\IEEEauthorblockA{Department of Electrical Engineering\\
Korea Advanced Institute of Science and Technology\\
Daejeon, 305-338, Republic of Korea\\
Email: jysohn1108@kaist.ac.kr, shyoon8@kaist.ac.kr, jmoon@kaist.edu }
}


%


\maketitle

\begin{abstract}

The pilot reuse issue in massive multi-input multi-output (MIMO) antenna systems
with interfering cells is closely examined.
This paper considers scenarios where 
the ratio of the channel coherence time to the number of users in a cell 
may be sufficiently large.
One such practical scenario arises when the number of users per unit coverage area cannot grow freely while user mobility is low, as in indoor networks.
Another important scenario is when the service provider is interested in 
maximizing the sum rate over a fixed, selected number of users rather than the sum rate over all users in the cell.
A sum-rate comparison analysis shows that in such scenarios
less aggressive reuse of pilots involving allocation
 of additional pilots for interfering users 
yields significant performance advantage relative to the case where all cells reuse the same pilot set.
For a given ratio of the normalized coherence time interval to the number of users per cell,
the optimal pilot assignment strategy is revealed via a closed-form solution
and the resulting net sum-rate is compared with that of the full pilot reuse.

 
\end{abstract}


%
\IEEEpeerreviewmaketitle

\section{Introduction}

The MIMO antenna technology has become an essential component of modern cellular wireless communication systems.
An exciting recent development related to MIMO, collectively  referred to as massive MIMO or a large-scale antenna array system,
is the potential deployment of a very large number of antenna elements at base stations (BSs).
The pioneering work of 
 \cite{ref:Marzetta2} has demonstrated that 
in time-division duplex (TDD) operation with uplink training for attaining channel state information (CSI), 
the effects of uncorrelated noise and fast fading disappear as the number of BS antennas increases without limit,
while only the effect of degraded channel estimation due to 
pilot contamination from the reuse of the same pilot set in interfering cells
limits the sum-rate performance.  

Systematic methods to mitigate pilot contamination effect have been studied by various researchers \cite{ref:Appaiah}\cite{ref:Yin}\cite{ref:Ashikhmin}. 
Shifting of pilot frames corresponding to neighboring cells was discussed in \cite{ref:Appaiah}, while \cite{ref:Yin} suggested coordinated pilot assignment using second order statistics for the channels. In \cite{ref:Ashikhmin}, efficient outer multi-cellular precoding under the assumption of appropriate cooperation among BSs was introduced for mitigating the pilot contamination effect. All known works related to combating pilot contamination deal with improving channel estimation and rarely mention the potential of allowing more orthogonal pilots. 

In Appendix F of \cite{ref:Marzetta2}, Marzetta briefly discusses the performance impact of serving a smaller number of terminals 
than the number of available orthogonal pilot patterns. He concludes that while the per-user signal-to-interference ratio (SIR) would improve due to reduced pilot contamination
(as some pilot patterns can be assigned to interferers), the per-cell throughput will suffer as the number of users has been reduced. This observation is based on the fact that the mean throughput per cell improves as the logarithm of SIR 
while it degrades more rapidly as a linear function of the decreasing number of users. 
Let $T_{coh}$ be the coherence time interval and $T_{del}$ be the channel delay spread. It is convenient to express the coherence time interval as a dimensionless quantity $N_{coh}=T_{coh}/T_{del}$, after normalization by the channel delay spread. Let $K$ denote the number of users in a cell. If the sum rate is predominantly a linear function of $K$, then
the net sum rate, after discounting the time allocated for the pilot, becomes proportional to $K(1-K/N_{coh})$ assuming $K$ orthogonal pilots are used to train $K$ channels. It is then easy to see that the net rate is maximized when $K=N_{coh}/2$. Marzetta's message on pilot reuse is basically that given a fixed $N_{coh}$, the sum rate will be maximized when $K$ is set to $N_{coh}/2$ and the corresponding $N_{coh}/2$ pilots should be fully reused across all cells; reducing $K$ below $N_{coh}/2$ in an attempt to reduce interference will only hurt the overall throughput.

This message is certainly correct but is not applicable to cases where $K$ cannot simply be adjusted
for a given $N_{coh}$. In particular, we are concerned with the scenarios where the ratio 
$N_{coh}/K$ is considerably larger than 2. Such a scenario
arises when user mobility is low and at the same time the number of users that can be accommodated in a cell cannot be increased arbitrarily, as in indoor networks. 
Another feasible scenario is when the service provider wishes to maximize the sum rate over a limited number of users (e.g., high-paying customers demanding guaranteed data rates) rather than over all users in the cell.
In fact, higher $N_{coh}/K$ ratios due to a relatively small number of users per cell
reflect meaningful situations in 5G in light of the trend toward smaller cells.

In this paper, we evaluate the maximum net sum rate for any given ratio of $N_{coh}/K$ and find the corresponding optimal pilot reuse strategy under some mild constraints on cell geometry 
and assuming an infinitely large BS antennas.
We show that as the ratio $N_{coh}/K$ increases, it is beneficial to actually prevent full pilot reuse across neighboring cells, i.e., it makes sense to allow additional pilot patterns to remove interference from users in the neighboring cells even at the expense of increased training time.
The optimal pilot assignment solution is in general not trivial and requires a careful mathematical derivation.

A simple example illustrates when it becomes sensible to depart from the full pilot reuse.
Using the well-known asymptotic result for a large number of BS antennas \cite{ref:Marzetta2} and accounting for the pilot portion of the coherent interval not used for the actual data transmission,
the net user information rate for the full pilot reuse (or a reuse factor of one) is 
given by  
\begin{equation}\label{equation:1}
C_{net,1}=(1-K/N_{coh})\log_2(1+SIR).
\end{equation}
Assume that the interference is coming from a single user located at the nearest vertex of a neighboring cell. The cells are hexagonal. 
For comparison, now assume a pilot reuse factor of three, where the entire cells are divided into three equal partitions, just as in the familiar case of the frequency reuse factor three. 
Again assume for simplicity that there is only one interferer who is located in the nearest vertex of the nearest cell using the common set of pilots. 
Given that only the cells in a given partition reuse the same pilot set, the distance of the worst case interferer from the BS
is doubled compared to the case for a reuse factor of one. This improves the SIR by the factor $(2^{\gamma})^2=4^{\gamma}$, where $\gamma$ represents the exponent of signal decay as a function of distance. On the other hand, the pilot interval is now increased by a factor of three to accommodate three distinct orthogonal pilot sets. Accordingly, the net user rate for the pilot reuse factor 3 is
\begin{equation}\label{equation:1}
C_{net,3}=(1-3K/N_{coh})\log_2(1+4^{\gamma}SIR).
\end{equation}
It is straightforward to show that $C_{net,3}$ becomes larger than $C_{net,1}$, when 
\begin{equation}
\frac{N_{coh}}{K} > 3 + \frac{\log_2 SIR}{\gamma}.
\end{equation}
With $\gamma = 3.8$ and an SIR of, say, 3.7 dB, we observe that 
as long as $N_{coh}/K > 6.2$, it is better to go with the pilot reuse factor 3 relative to the full reuse.

The exact threshold value for $N_{coh}/K$
for which less than full pilot reuse becomes beneficial will change depending on the assumed locations of the user and interferers as well as the cell partition structure, 
but it remains true that for given SIR, coherence time and number of users in a cell, there exists an optimal value of 
the pilot reuse factor that is not equal to one in general.
We also show that there exists an optimal way of partitioning cells in terms of pilot reuse; 
we provide a closed-form solution to the problem of optimal partitioning wherein 
only the cells in a given partition reuse the same set of pilots.
 
This paper is organized as follows. Section II gives an overview of the system model for massive MIMO and the 
pilot contamination effect.
Section III discusses the assumptions made 
on cell geometry and basic partitioning steps needed to utilize
longer pilots, and presents the mathematical analysis for specifying the optimal pilot assignment strategy for the case of single user per cell ($K=1$). 
Simulation results are also presented in this section. 
In Section IV, the solutions are extended to the general case with multiple users per cell. Section V draws conclusions.
\section{Pilot Contamination Effect for Massive MIMO Multi-Cellular System}

\subsection{System Model}
We assume that the network consists of $L$ hexagonal cells with $K$ users per cell who are uniform-randomly located.
Downlink CSIs are estimated at each BS by uplink pilot training assuming channel reciprocity in TDD operation. 
The channel model of \cite{ref:Marzetta2} is assumed in this paper. The complex propagation coefficient $g$ of a link can be decomposed into a complex fast fading factor $h$ and a slow fading factor $\beta$. Therefore, the channel between the $m^{th}$ BS antenna of the $j^{th}$ cell and the $k^{th}$ user of the $l^{th}$ cell are modeled as:
\begin{equation} \label{channel model}
g_{mjkl}=h_{mjkl}\sqrt{\beta_{jkl}}.
\end{equation}
The slow fading factor, which accounts for the geometric attenuation and shadow fading, is modeled as
\begin{equation} \label{slow fading model}
\beta_{jkl}=\frac{z_{jkl}}{r_{jkl}^\gamma}
\end{equation}
where $r_{jkl}$ is the distance between the $k^{th}$ user in the $l^{th}$ cell and the base station in the $j^{th}$ cell. The parameter $\gamma$ represents the signal decay exponent, 
while $z_{jkl}$ is a log-normal random variable.


\subsection{Pilot Contamination Effect}
The pilot contamination effect is the most serious issue that arises in multi-cell TDD systems with very large BS antenna arrays. 
For uplink training, each BS collects pilot sequences sent by its users. Usually, orthogonal pilot sequences are assigned to users in a cell so that the channel estimate for each user 
does not suffer from interferences from other users in the same cell. However, the use of the same pilot sequences for users in other cells cause the channel estimates to be contaminated, 
and this effect, called pilot contamination, limits the achievable rate, as $M$, the number of BS antennas, tends to infinity. 

According to \cite{ref:Marzetta2}, under the assumption of a single user per cell, the achievable rate during downlink data transmission for the user in the $j^{th}$ cell contaminated by users with the same pilot on other cells is given for a large $M$ by
\begin{equation} \label{achievableR}
\log_{2} \left(1+\frac{\beta_{jj}^{2}}{\sum_{l\neq j}\beta_{jl}^{2}}\right)
\end{equation}
where $\beta_{jl}$ is the slow fading component of the channel between the  $j^{th}$ BS and the interfering user in the $l^{th}$ cell. In the limit of large $M$, the achievable rate depends only on the ratio of the signal to interference due to the pilot reuse. 


\section{Pilot Assignment Strategy for Single User Case}
In this section, we provide analysis on how much time should be allocated for channel training given a coherence time and how the pilot sequences should be assigned to users on multiple cells.
We derive optimal pilot assignment strategy, which mitigates the pilot contamination effect and maximizes total bits transmitted in a given $T_{coh}$. 
Our analysis considers using pilot sequences possibly longer than the number of users in each cell, while orthogonality of the pilots within a cell is guaranteed. 
For simplicity, all cells are restricted to have only one user, i.e, $K=1$ in this section. The results are extended to $K>1$ in the next section.


\subsection{Hexagonal-Lattice Based Cell Clustering}

\begin{figure}[!t]
    \centering
    \includegraphics[height=45mm]{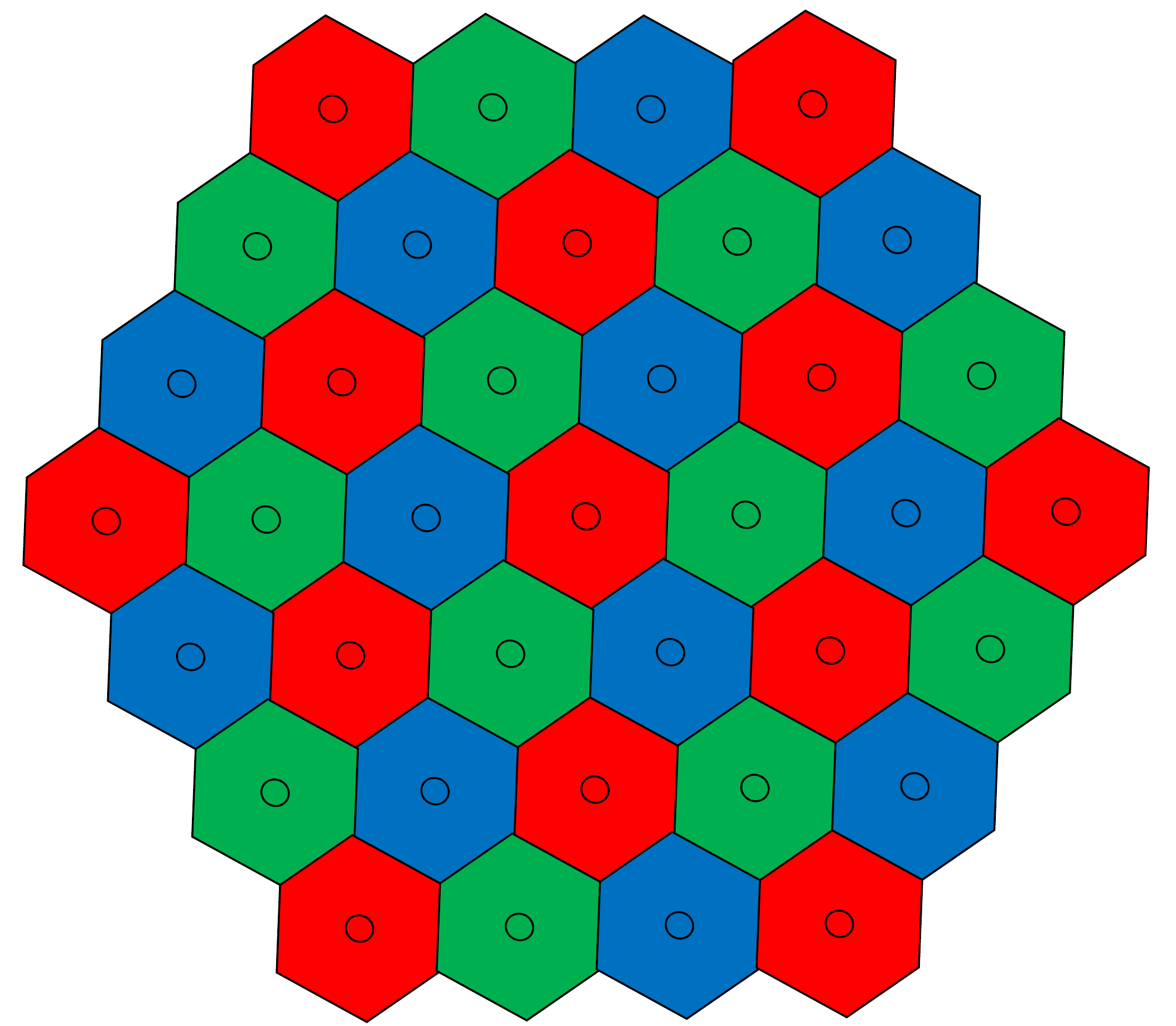}
    \caption{3-way Partitioning}
    \label{Fig:3-lattice Partitioning}
\end{figure}

Consider $L$ hexagonal cells. Imagine partitioning these cells into three equi-distance subsets 
maintaining the same lattice structure as depicted in Fig. \ref{Fig:3-lattice Partitioning}. 
This partitioning is identical to the familiar partitioning of contiguous hexagonal cells for utilizing three frequency bands according to a frequency reuse factor of three.

    

\begin{figure}[!t]
\centering
        \includegraphics[height=45mm]{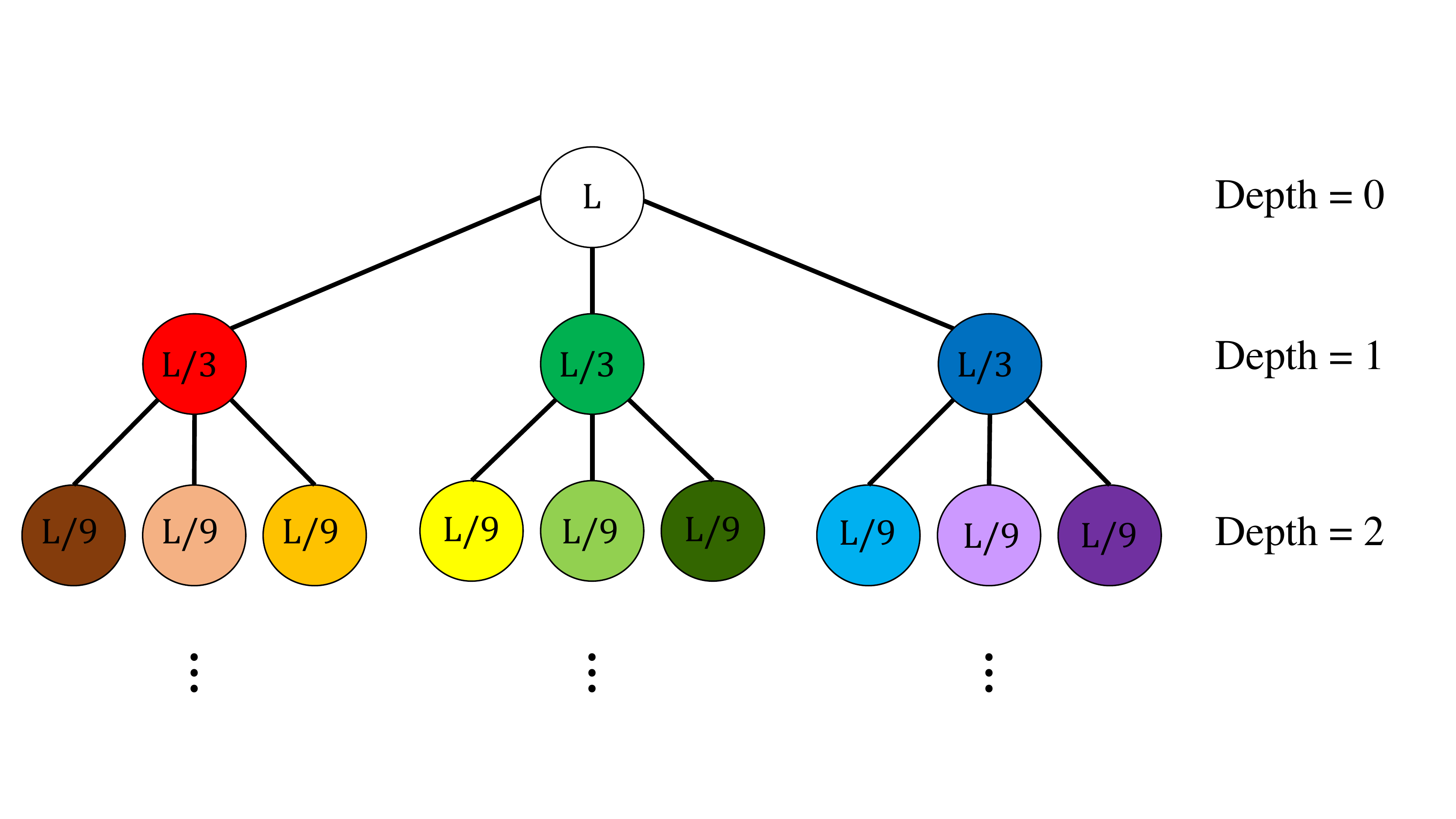}
    \caption{Hierarchical set partitioning}
    \label{Fig:Hierarchy tree structure}
\end{figure}

It can easily be seen that each coset, having the same hexagonal lattice structure, can be further partitioned in the similar way.
The partitioning can clearly be applied in a successive fashion, giving rise to the possibility of hierarchical set partitioning of the entire cells.
In the tree structure of Fig.\ref{Fig:Hierarchy tree structure}, the root node represents the original group of  $L$ contiguous hexagonal cells,
and the three children labeled $L/3$ corresponding to the three colored-cosets of Fig. \ref{Fig:3-lattice Partitioning}. Also, applying a 3-way partitioning to a coset results in additional three child nodes with labels $L/9$. Note that a node at depth $i$ corresponds to a subset of $L3^{-i}$ cells. A coloring scheme is also employed to 
represent the same pilot sets with  common color. The number of leaves with different (non-white) colors represents the number of different orthogonal pilot sets used.
A leaf (end node) with a single cell would correspond to a depth of $\log_{3}L$, but we do not allow such leaves in our analysis as this means
there would be users with no pilot contamination, thus driving the average per-cell throughput of the network to infinity 
as $M$ grows. This particular situation would not lend itself to a meaningful mathematical analysis. Thus, the maximum depth of a leaf in our tree is set to $\log_3 L -1$. 

\subsection{Pilot Assignment Vector}

\begin{figure}[!t]

    \includegraphics[height=45mm]{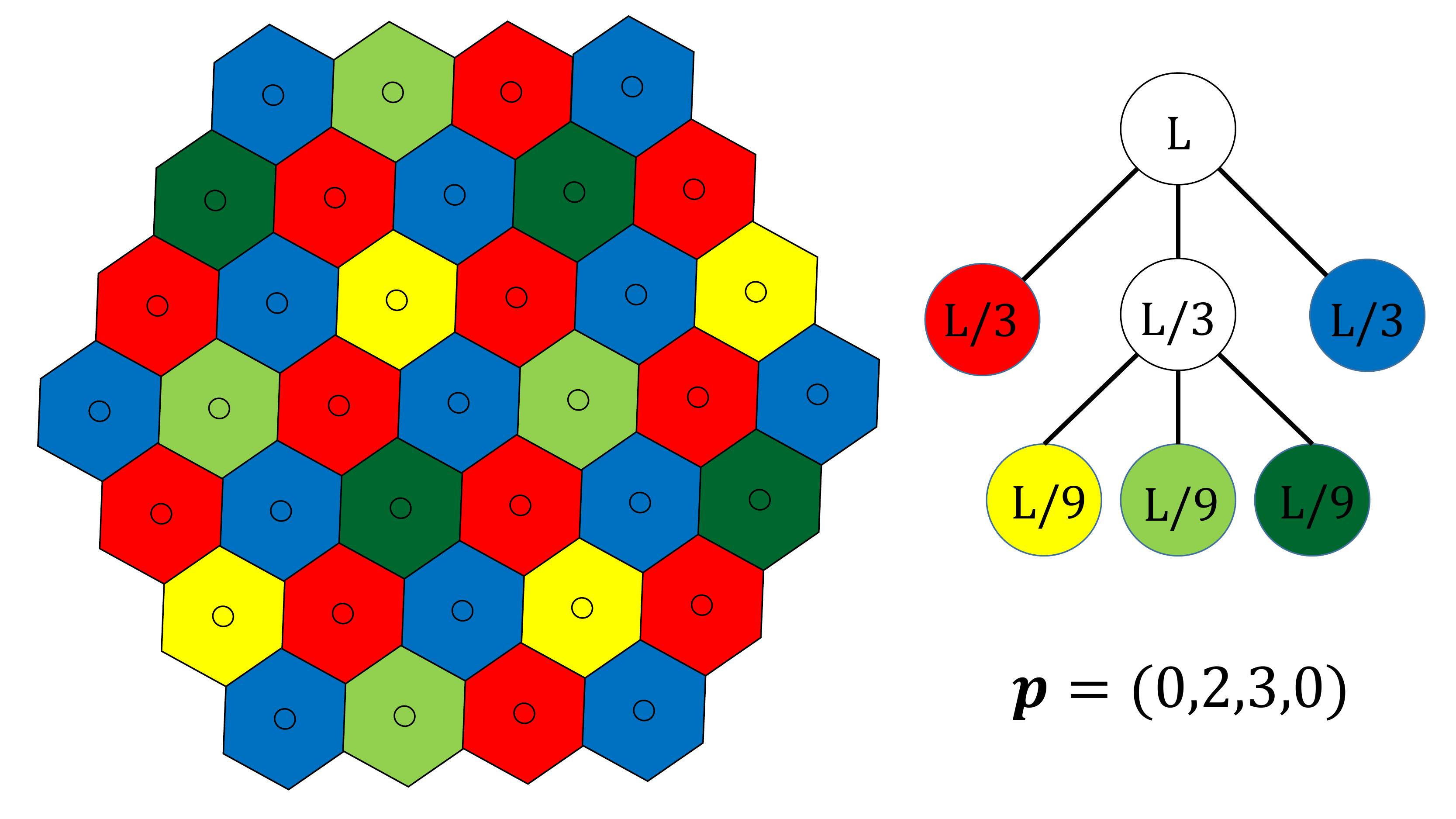}
    \caption{Example of coloring, tree structure, and pilot assignment vector}
    \label{Fig:coloring example}
    
\end{figure}

The pilot assignment method can conveniently be formulated in a vector form. Let $\mathbf{p}$ be a vector with element 
$p_{i}$, $i=0,\cdots,\log_3 L -1$, representing the number of leaves at depth $i$ of the partitioning tree. For example, 
for the tree of Fig. \ref{Fig:coloring example}, we have $\mathbf{p}=(0,2,3,0)$, as there are two leaves at depth 1 and three at depth 2.

\newtheorem{theorem}{Theorem}
\newtheorem{lemma}{Lemma}
\newtheorem{corollary}{Corollary}
\newenvironment{definition}[1][Definition]{\begin{trivlist}
\item[\hskip \labelsep \normalfont #1]}{\end{trivlist}}

\begin{definition}[Definition:\nopunct]
For the given $L$ cells with $K=1$, $P_{L}$ is a set of valid pilot assignment vectors based on 3-way partitioning and is given by
\begin{eqnarray} \label{pilot assgning vector}
P_{L} = \{\mathbf{p}=(p_0,p_1,\cdots,p_{\log_3 L-1})\:\:|\enspace 0 \leq p_i \leq 3^i 
\nonumber\\
p_{i} \textrm{: integer and} \sum\limits_{i=0}^{\log_3 L -1} p_i3^{-i}=1\}
\end{eqnarray}
\end{definition}

\begin{definition}[Definition:\nopunct]
For the given $L$ cells, $N_{pil}(\mathbf{p})$ is the length of the valid pilot assignment vector $\mathbf{p}=(p_0,p_1,\cdots,p_{\log_3 L -1})$ and is given by
\begin{eqnarray} \label{pilot length function}
N_{pil}(\mathbf{p}) = \sum\limits_{i=0}^{\log_3 L-1} p_i
\end{eqnarray}
\end{definition}
The pilot length $N_{pil}(\mathbf{p})$ represents the number of orthogonal pilots corresponding to the given pilot assignment vector $\mathbf{p}$.
As an example, for the pilot assignment strategy shown in Fig. \ref{Fig:coloring example}, we have $N_{pil}(\mathbf{p})=5$.

Notice that the users in different cells belonging to a given leaf experience pilot contamination. The severity of the contamination depends on 
the depth of the leaf. Every time the depth is increased, the distance between interfering cells (cells that reuse the pilot set)
increases by a certain factor, as can be observed from Fig. \ref{Fig:coloring example}. In fact, the distance grows geometrically as the depth increases.
According to (\ref{achievableR}), the achievable rate is determined by the $\beta$ values of the interfering users, which in turn depend on 
the distances of the interferers from the BS. 
The distance growth manifests itself as the reduced pilot contamination effect or an improved SIR, increasing the throughput. 
More specifically, the throughput grows roughly with $\log_2(\alpha^{2\gamma} SIR_1)=2\gamma \log_2(\alpha)+\log_2 SIR_1$, where 
$SIR_1$ is the reference SIR corresponding to the pilot reuse factor 1 and $\alpha$ is the parameter that represents the distance growth. It is clear that a geometric growth of 
$\alpha$ gives rise to a linear increase in the throughput. 
Letting $C_{i}$ be the rate of a user at depth $i$, this is to say that $C_{i}$ increase linearly with depth $i$.

The per-cell sum rate for the network can be expressed as 
\begin{equation} \label{sum rate}
C_{sum}(\mathbf{p})=\frac{1}{L}\sum\limits_{i=0}^{\log_3 L-1} L3^{-i} p_i  C_{i}=\sum\limits_{i=0}^{\log_3 L-1} 3^{-i} p_i  C_{i}.
\end{equation}
The per-cell net sum rate, accounting for the fact that useful data gets transmitted only over the portion of the coherence time not allocated to the pilots,
is given by 
\begin{align} \label{net sum rate}
C_{net}(\mathbf{p},N_{coh}) &=\frac{[N_{coh}-N_{pil}(\mathbf{p})]}{N_{coh}} C_{sum}(\mathbf{p}).
\end{align}
We shall use $C_{net}$ as the objective function for 
finding optimal pilot assignment strategies.

\subsection{Closed-Form Solution of Optimal Pilot Assignment Strategy}
The optimal pilot assignment vector $\mathbf{p}_{opt}(N_{coh})$ for the given normalized coherence time $N_{coh}$ and the number of cells $L$ is:
\begin{equation}\label{optimal assigning}
\mathbf{p}_{opt}(N_{coh}) = \underset{\mathbf{p}\in P_{L}}{\arg\max}\ C_{net}(\mathbf{p},N_{coh}).
\end{equation}
We also note that for $K=1$, all valid pilot assignments yield odd pilot lengths, as formally stated in Lemma \ref{Lemma_pilot_length}.

\begin{lemma} \label{Lemma_pilot_length}
For $K=1$ and given $L$, $\{N_{pil}(\mathbf{p}) : \mathbf{p} \in P_{L}\} = \{1,3,5,\cdots, \frac{L}{3} \}$.
\end{lemma}


Before giving the first main theorem, it is useful to define the pilot assignment vector that maximizes the per-cell sum rate $C_{sum}$ with a finite pilot length constraint:
\begin{equation}\label{optimal_prime}
\mathbf{p}'_{opt}(N_{p0})=\underset{\mathbf{p}\in \Omega(N_{p0})}{\arg\max} \ C_{sum}(\mathbf{p})
\end{equation}
where 
\begin{equation}\label{Omega_variable}
\Omega(N_{p0})=\{\mathbf{p}\in P_{L} \: | \: N_{pil}(\mathbf{p})=N_{p0}\}.
\end{equation}
We will first find a closed-form solution for $\mathbf{p}'_{opt}$ and then 
find eventually $\mathbf{p}_{opt}$ by exploring its relationship with the former. 

\begin{definition}[Definition:\nopunct]

For each valid pilot assignment vector $\mathbf{p}=(p_{0},p_{1},\cdots,p_{\log_3 L-1})$, the corresponding transition vector $\mathbf{t}=(t_{0},t_{1},\cdots,t_{\log_3 L-2})$ is defined as:
\begin{equation}\label{ptot}
\begin{cases}
t_{0} = 1 - p_{0} \\
t_{i} = -p_{i} + 3t_{i-1}. & 1\leq i \leq  \log_3 L-2 \\
\end{cases}
\end{equation}
The inverse relationship exists:
\begin{equation}\label{ttop}
\begin{cases}
p_{0} = 1 - t_{0} \\
p_{i} = 3t_{i-1}-t_{i} & 1\leq i \leq  \log_3 L-2 \\
p_{\log_3 L-1}=3t_{\log_3 L-2}.
\end{cases}
\end{equation}

\end{definition}
The first two equations of (\ref{ttop}) come from (\ref{ptot}), while the last equation is from the fact $\mathbf{p},     \sum_{i=0}^{\log_3 L-1}p_{i}3^{-i}=1$.
Each element of $\mathbf{t}$ represents the number of transitions at depth $i$ from the conventional (i.e., full pilot reuse) assignment $(1,0,\cdots,0)$. 
For example, $\mathbf{p}=(1,0,0,0)$ turns to $\mathbf{p}=(0,2,3,0)$ via a transition vector $\mathbf{t}=(1,1,0)$. The first transition element
$t_0=1$ triggers a (3-way) partitioning at depth 0, temporarily creating a pilot vector (0,3,0,0). The next element $t_1=1$ induces a (3-way) partitioning on one of the 
3 existing partitions at depth 1, thereby giving rise to a new pilot vector (0,2,3,0). Since the next transition vector element is zero, the partitioning stops.
The transition elements also point to the number of white nodes at each depth, 
as can be confirmed in Fig. \ref{Fig:coloring example}.

\begin{lemma} \label{Lemma2}
Any transition vector $\mathbf{t}=(t_{0},t_{1},\cdots,t_{\log_3 L-2})$ originated from $\mathbf{p} \in \Omega(N_{p0})$ satisfies
\begin{equation} \label{Lemma2 result}
\begin{cases}
0 \leq t_{i} \leq 3^{i} & 0 \leq i \leq \log_3 L-2 \\
\displaystyle\sum_{i=0}^{\log_3 L-2}t_{i}=\frac{N_{p0}-1}{2}. 
\end{cases}
\end{equation}
\end{lemma}
We further define the index function  $\chi(N_{p0})$ that identifies the first non-zero position of $\mathbf{p}'_{opt}(N_{p0})$: 
\begin{equation}\label{xi_index}
\chi(N_{p0})=\min\{k \: | \: \sum_{i=0}^{k}3^{i}>\frac{N_{p0}-1}{2} \}
\end{equation}

\begin{theorem} \label{Theorem1}
For $K=1$, given $L$ and $N_{p0}$, the optimal pilot assignment vector $\mathbf{p}'_{opt}(N_{p0})=(p'_{0},\cdots,p'_{\log_3 L -1})$ with respect to $C_{sum}$, has its components as follows:
\begin{equation} \label{Thm1 result}
p'_{i} =
\begin{cases}
\displaystyle\sum_{s=0}^{i}3^{s}-\frac{N_{p0}-1}{2} & i=\chi(N_{p0}) \\
3\left(\displaystyle\frac{N_{p0}-1}{2}-\displaystyle\sum_{s=0}^{i-2}3^{s}  \right) & i=\chi(N_{p0})+1 \\
0 & \textrm{otherwise}
\end{cases}
\end{equation}
\end{theorem}
Here we only provide a brief sketch of the proof due to space limitation. The sum rate in (\ref{sum rate}) can be expressed using the transition vector $\mathbf{t}$ via (\ref{ptot}):
\begin{equation}
C_{sum}(\mathbf{t})= C_{0} + \sum\limits_{i=0}^{\log_3 L-2}t_{i}3^{-i}(C_{i+1}-C_{i}).
\end{equation}
Because $C_{i}$ is a linear function of $i$, the difference $(C_{i+1}-C_{i})$ is a constant. Therefore, all we have to do is to find the optimal $\mathbf{t}$ which maximizes $\sum_{i=0}^{\log_3 L-2}t_{i}3^{-i}$.

Under the condition of Lemma \ref{Lemma2}, for maximizing $\sum_{i=0}^{\log_3 L-2}t_{i}3^{-i}$, we should give the largest values $t_{i}=3^{i}$ to the lower indices $i<\chi(N_{p0})$ and the remaining values to the rest of the indices. The final solution for $\mathbf{t}'=(t_{0}',t_{1}',\cdots,t_{\log_3 L-2}')$ is:
\begin{equation} \label{t_i'}
t_{i}'=
\begin{cases}
3^{i} & i < \chi(N_{p0}) \\
\frac{N_{p0}-1}{2} - \displaystyle\sum_{i=0}^{\chi(N_{p0}) - 1}3^{i}	& i= \chi(N_{p0}) \\
0 & i > \chi(N_{p0}),
\end{cases}
\end{equation}
which leads to (\ref{Thm1 result}) via (\ref{ttop}).

For a given $N_{p0}$, more than one valid pilot assignments may exist. For example, if $L=81$ and $N_{p0}=7$, $\mathbf{p}=(0,1,6,0)$ and $\mathbf{p}=(0,2,2,3)$ are valid vectors, but Theorem \ref{Theorem1} reveals that $\mathbf{p}'_{opt}(N_{p0})=(0,1,6,0)$.

From Theorem \ref{Theorem1}, a certain trend relating the optimal pilot assignment vectors $\mathbf{p}'_{opt}(N_{p0})$ and  $\mathbf{p}'_{opt}(N_{p0}+2)$ can be observed as follows.

\begin{corollary} \label{Corollary1}
For two pilot lengths $N_{p0}$ and $N_{p0}+2$, the two corresponding optimal pilot assignment vectors $\mathbf{p}'_{opt}(N_{p0})=(p^{*}_{0},\cdots,p^{*}_{\log_3 L -1})$ and $\mathbf{p}'_{opt}(N_{p0}+2)=(p^{**}_{0},\cdots,p^{**}_{\log_3 L -1})$ exhibit the following relationship:
\begin{equation}
p^{**}_{i} = 
\begin{cases}
p^{*}_{i}-1 & i=\chi(N_{p0}) \\
p^{*}_{i}+3 & i=\chi(N_{p0})+1 \\
p^{*}_{i} & \textrm{otherwise}
\end{cases}
\end{equation}
\end{corollary}

For a given $N_{p0}$, the optimal assignment vectors $\mathbf{p}'_{opt}(N_{p0})$ and $\mathbf{p}'_{opt}(N_{p0}+2)$ show a predictable pattern of tossing 1 from the left-most non-zero component to increase the adjacent component by 3. For example, in the case of $L=81$, the optimal assignment $\mathbf{p}'_{opt}(7)=(0,1,6,0)$ can be transformed by reducing the second component by 1 and increasing the third one by 3, which results in the next optimal assignment for $N_{p0}=9$: $\mathbf{p}'_{opt}(9)=(0,0,9,0)$. It can be seen that there is a tendency to reduce the left-most non-zero values which give the most severe pilot contamination.

We now set out to find $\mathbf{p}_{opt}(N_{coh})$.
Theorem \ref{Theorem1} and Corollary $\ref{Corollary1}$ already identify, given the fairly mild constraints of 
hexagonal cells and equi-distance partitioning, the optimal pilot assignment strategy maximizing the sum rate
for the chosen pilot sequence length. The next step is to find the relationship between the normalized coherence time $N_{coh}$ and the optimal pilot sequence length.

First, write the net sum-rate as 
\begin{gather} \label{eq24}
C_{net}(\mathbf{p}'_{opt}(N_{p0}),N_{coh}) \nonumber\\
=\frac{(N_{coh}-N_{p0})}{N_{coh}}  C_{sum}(\mathbf{p}'_{opt}(N_{p0}))
\end{gather}
which is an increasing function of $N_{coh}$ and crosses the horizontal axis once at $N_{coh}=N_{p0}$. Moreover this function saturates to $C_{sum}(\mathbf{p}'_{opt}(N_{p0}))$ for very large $N_{coh}$. Imagine plotting this function 
for all possible values of $N_{p0}=1,3,5,7,\cdots$.
As $N_{p0}$ increases, the zero-crossing is naturally shifted to the right while the saturation value moves up. More specifically, the $C_{net}$ curve for $N_{p0}=2n-1$ and 
that for $N_{coh}=2n+1$ intersect once. On the left side of this intersection point,
the $C_{net}$ curve for $N_{p0}=2n-1$ is above the latter while on the right side, the latter curve is higher than the former. Let the horizontal value of this intersection point be 
$N_{coh}=\Delta_{n}$. Applying the same argument, the two $C_{net}$ curves for $N_{p0}=2n+1$ and $N_{p0}=2n+3$ intersect at $\Delta_{n+1}$ somewhere to the right of the previous intersection point 
$\Delta_{n}$. In between $\Delta_{n}$ and $\Delta_{n+1}$, the $C_{net}$ curve for 
$N_{coh}=2n+1$ yields the highest values, indicating that when $N_{coh}$ falls between the two intersection points, the optimal pilot length is $2n+1$. It can be shown that the intersection points are given by
\begin{equation}\label{delta_n}
\Delta_{n} = 
2\left( 2n-1-\displaystyle \sum_{i=0}^{\chi(2n-1)-1} 3^{i}+\xi(n) \right)+1,  \quad   1\leq n\leq N_{L}\\
\end{equation} 
where $\xi(n)=3^{\chi(2n-1)} C_{\chi(2n-1)}/(C_{\chi(2n-1)+1}-C_{\chi(2n-1)})$, with $C_{i}$ already defined in Section III-B, and $N_{L}$ is the number of all possible pilot lengths minus one.
Since $N_{pil}=1,3,\cdots,L/3$ from Lemma \ref{Lemma_pilot_length}, we have $N_{L}=\frac{1}{2}(\frac{L}{3} - 1)$.

We now state our second main theorem without a formal proof (which will be given elsewhere).
\begin{theorem} \label{Theorem2}
For $K=1$, given $L$ and $N_{coh}$, if the normalized coherent time $N_{coh}$ is in between two adjacent time points $\Delta_{n}$ and $\Delta_{n+1}$, i.e., $\Delta_{n}\leq N_{coh} <\Delta_{n+1}$, then the optimal assignment vector $\mathbf{p}_{opt}(N_{coh})$ satisfies
\begin{gather}
N_{pil}\left(\mathbf{p}_{opt}(N_{coh})\right)=2n+1 \nonumber\\
\mathbf{p}_{opt}(N_{coh})=\mathbf{p}'_{opt}(2n+1), \label{Thm2 result}
\end{gather}
where $\mathbf{p}'_{opt}(2n+1)$ is the vector maximizing $C_{sum}$ given the fixed pilot length ($2n+1$).
\end{theorem}

\subsection{Simulation Results}
Simulation is necessary to compute the $C_i$ values in (\ref{sum rate}). To compute $C_i$, the $\beta$ terms in (\ref{achievableR}) need to be generated pseudo-randomly according to the assumed underlying statistical properties.
For this, we assume: a signal decay exponent of $\gamma = 3.8$, the shadow-fading standard deviation of $\sigma_{shadow}=8.0$ dB, and a cell radius of $r_{c}=1600$ meters with a cell-hole radius of $r_{h}=100$ meters, to be consistent with the parameter values used in \cite{ref:Marzetta2}. The user locations are uniform-random. 
To generate each $C_i$ value, an average is taken over 100,000 pseudo-random trials.
The number of cells are fixed to $L=81$ and the number of users in each cell in this section is $K=1$.

\begin{table}
\caption{Optimal pilot assignment ($L=81,K=1$)}
\centering
\label{Table:Optimal assignment}
\begin{tabular}{|c|c|c|}
\hline
$N_{coh}$ & $\mathbf{p}_{opt}(N_{coh})$  & $N_{pil}(\mathbf{p}_{opt}(N_{coh}))$\tabularnewline
\hline
$0\sim 6$ & $(1,0,0,0)$ & $1$\tabularnewline
$7\sim 23$ & $(0,3,0,0)$ & $3$\tabularnewline
$24\sim 27$ & $(0,2,3,0)$ & $5$\tabularnewline
$28\sim 31$ & $(0,1,6,0)$ & $7$\tabularnewline
$32\sim 70$ & $(0,0,9,0)$ & $9$\tabularnewline
$71\sim 74$ & $(0,0,8,3)$ & $11$\tabularnewline
$75\sim 78$ & $(0,0,7,6)$ & $13$\tabularnewline
$79\sim 82$ & $(0,0,6,9)$ & $15$\tabularnewline
$83\sim 86$ & $(0,0,5,12)$ & $17$\tabularnewline
$87\sim 90$ & $(0,0,4,15)$ & $19$\tabularnewline
$91\sim 94$ & $(0,0,3,18)$ & $21$\tabularnewline
$95\sim 98$ & $(0,0,2,21)$ & $23$\tabularnewline
$99\sim 102$ & $(0,0,1,24)$ & $25$\tabularnewline
$103\sim$ & $(0,0,0,27)$ & $27$\tabularnewline
\hline
\end{tabular}
\end{table}

Once the $C_i$ values are computed, the optimal pilot length and the pilot assignment vector as well as the net throughput can be obtained for various coherence intervals $N_{coh}$. Table \ref{Table:Optimal assignment} shows the optimal pilot assignment results for various values of $N_{coh}$.
We confirm that the simulation results of Table \ref{Table:Optimal assignment} are consistent with 
the mathematical analysis given in Theorem \ref{Theorem2}.

\begin{figure}[!t]
\centering
        \includegraphics[height=60mm]{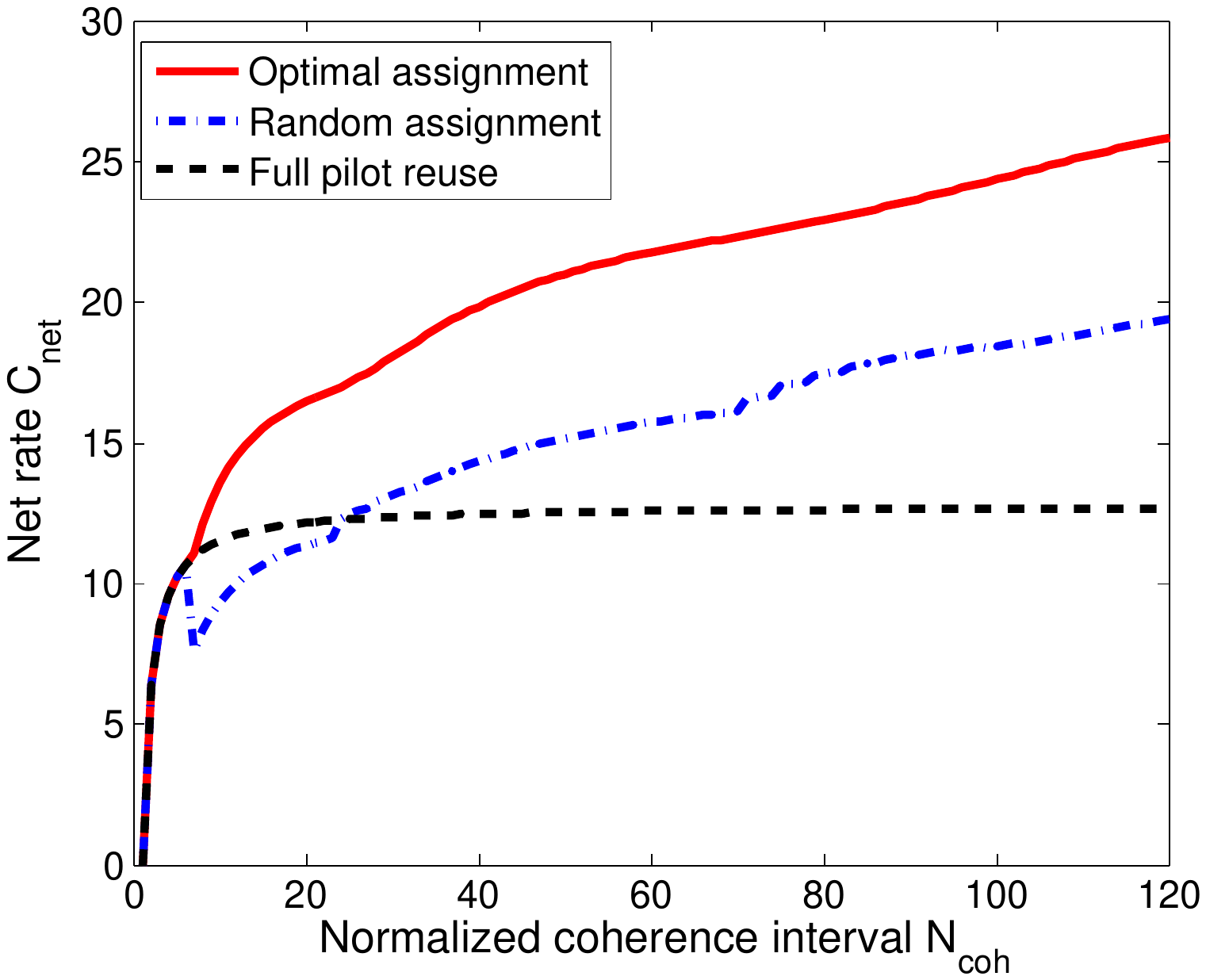}
    \caption{Net rate for various pilot assignments ($L=81,K=1$)}
    \label{Fig:K=1 Simulation}
    
\end{figure}

Fig. \ref{Fig:K=1 Simulation} shows the average achievable net rates for various pilot assignments versus normalized coherence interval. The random assignment means that a pilot sequence is chosen 
randomly and independently from $N_{pil}(\mathbf{p}_{opt}(N_{coh}))$ orthogonal pilots, and assigned to each user.
Therefore, the optimal assignment and the random assignment use the same amount of pilots for any given $N_{coh}$.
It can be seen that a substantial performance gain is obtained using the optimal method compared to the full pilot reuse case as $N_{coh}$ increases 
beyond 7. The random assignment is worse than the full reuse initially but eventually outperforms the latter as $N_{coh}$ grows.

For $N_{coh}=20,40$ and $80$, for example, the optimal assignment method has $35\%,59\%$ and $81\%$ higher net rates $C_{net}$ than the full pilot reuse assignment, respectively.
As coherence time increases, the benefit of allocating more time for pilots is considerable. 
Also, the non-shrinking performance gap between the optimal assignment and the random assignment indicate that
structured optimal assignment is required for a given pilot time, in order to maximize the net throughput of the network.

\section{Multi-User Analysis}
We now consider multiple users per cell. Due to space limitations we will state our main results 
in the form of two main theorems without proof.
Interestingly, the optimal pilot assignment 
vectors take a similar form to those for the  $K=1$ case.

\subsection{Analysis for Multi-User Case}
The set of valid pilot assignment vectors generalized to $K>1$ is defined as: 
\begin{eqnarray} \label{pilot assgning vector}
P_{L,K} = \{\mathbf{p}=(p_0,p_1,\cdots,p_{\log_3 L -1})\:\:|\enspace 0 \leq p_i \leq K3^i 
\nonumber\\
p_{i} \textrm{: integer and} \: \sum\limits_{i=0}^{\log_3 L -1} p_i3^{-i}=K \}
\end{eqnarray}
Also, generalize the following definitions:
\begin{equation}
\mathbf{p}_{opt}(N_{coh},K) = \underset{\mathbf{p}\in P_{L,K}}{\arg\max}\ C_{net}(\mathbf{p},N_{coh})
\end{equation}
\begin{equation}
\Omega(N_{p0},K)=\{\mathbf{p}\in P_{L,K} \: | \: N_{pil}(\mathbf{p})=N_{p0}\}
\end{equation}
\begin{equation}
\mathbf{p}'_{opt}(N_{p0},K)=\underset{\mathbf{p}\in \Omega(N_{p0},K)}{\arg\max} \ C_{sum}(\mathbf{p})
\end{equation}
\begin{equation}
\chi(N_{p0},K)=\min\{k \: | \: \sum_{i=0}^{k}K3^{i}>\frac{N_{p0}-K}{2} \}
\end{equation}
We present the following theorem for general $K$:
\begin{theorem} \label{Theorem3}
For given $N_{p0},L$ and $K$, the optimal pilot assignment vector $\mathbf{p}'_{opt}(N_{p0},K)=(p'_{0},\cdots,p'_{\log_3 L -1})$ maximizing $C_{sum}$ has its elements $p'_{i}$ in the form:
\begin{equation} \label{Thm3 result}
p'_{i} =
\begin{cases}
\displaystyle\sum_{t=0}^{i}K3^{t}-\frac{N_{p0}-K}{2} & i=\chi(N_{p0},K) \\
3\left(\displaystyle\frac{N_{p0}-K}{2}-\displaystyle\sum_{t=0}^{i-2}K3^{t}  \right) & i=\chi(N_{p0},K)+1 \\
0 & \textrm{otherwise}
\end{cases}
\end{equation}
\end{theorem}

We have another generalized definition:
\begin{definition}[Definition:\nopunct]
\begin{equation}\label{delta_n_K}
\Delta^{(K)}_{n} = \frac{2\left( 2n-1-\displaystyle\sum_{i=0}^{\eta(n,K)-1}K3^{i} + K\xi(n,K) \right)+K}{K}
\end{equation}
for $1\leq n\leq N_{K,L}$, where 
\begin{equation}
\begin{cases}
\eta(n,K)=\chi(2n+K-2,K) \\
\xi(n,K)=3^{\eta(n,K)}\frac{C_{\eta(n,K)}}{C_{\eta(n,K)+1}-C_{\eta(n,K)}}\\
N_{K,L}=\frac{\frac{LK}{3} - K}{2}
\end{cases}
\end{equation}
At the initial point, $\Delta^{(K)}_{0}=0$.
\end{definition}
Using (\ref{delta_n_K}), Theorem \ref{Theorem2} can be generalized as follows:
\begin{theorem} \label{Theorem4}
For given $L$,$K$ and $N_{coh}$, if the ratio $N_{coh}/K$ is in between two adjacent time points $\Delta^{(K)}_{n}$ and $\Delta^{(K)}_{n+1}$, i.e., $\Delta^{(K)}_{n}\leq \frac{N_{coh}}{K} <\Delta^{(K)}_{n+1}$, then the optimal assignment vector $\mathbf{p}_{opt}(N_{coh},K)$ maximizing $C_{net}(\mathbf{p},N_{coh})$ satisfies
\begin{gather}
N_{pil}\left(\mathbf{p}_{opt}(N_{coh},K)\right)=2n+K \nonumber\\
\mathbf{p}_{opt}(N_{coh},K)=\mathbf{p}'_{opt}(2n+K,K), \label{Thm2 result}
\end{gather}
where $\mathbf{p}'_{opt}(2n+K,K)$ is the vector for fixed pilot length ($2n+K$) that maximizes $C_{sum}$ for given $L$ and $K$.
\end{theorem}

\subsection{Simulation Result}

In Table \ref{Table:Optimal assignment2}, the optimal assignment vectors and pilot lengths for various $N_{coh}$ are shown, assuming $L=81$ and $K=2$.
Like in the case for $K=1$, the optimal pilot assignment vectors have a predictable form. As $N_{coh}$ increases, the optimal pilot sequence gradually becomes longer. Moreover, the optimal assignment vectors show a pattern of tossing 1 from the left most non-zero component to increase the adjacent component by 3, for an example $(0,6,0,0) \rightarrow (0,5,3,0)$ (consistent with Corollary $\ref{Corollary1}$).

\begin{table}
\caption{Optimal pilot assignment information ($L=81,K=2$)}
\centering
\label{Table:Optimal assignment2}
\begin{tabular}{|c|c|c|}
\hline
$N_{coh}$ & $\mathbf{p}_{opt}(N_{coh},K)$  & $N_{pil}(\mathbf{p}_{opt}(N_{coh},K))$\tabularnewline
\hline
$3\sim 11$ & $(2,0,0,0)$ & $2$\tabularnewline
$12\sim 15$ & $(1,3,0,0)$ & $4$\tabularnewline
$16\sim 45$ & $(0,6,0,0)$ & $6$\tabularnewline
$46\sim 49$ & $(0,5,3,0)$ & $8$\tabularnewline
$50\sim 53$ & $(0,4,6,0)$ & $10$\tabularnewline
$\vdots$ & $\vdots$ & $\vdots$\tabularnewline
$208\sim $ & $(0,0,0,54)$ & $54$\tabularnewline
\hline
\end{tabular}
\end{table}



    


\begin{figure}[!t]
\centering
        \includegraphics[height=60mm]{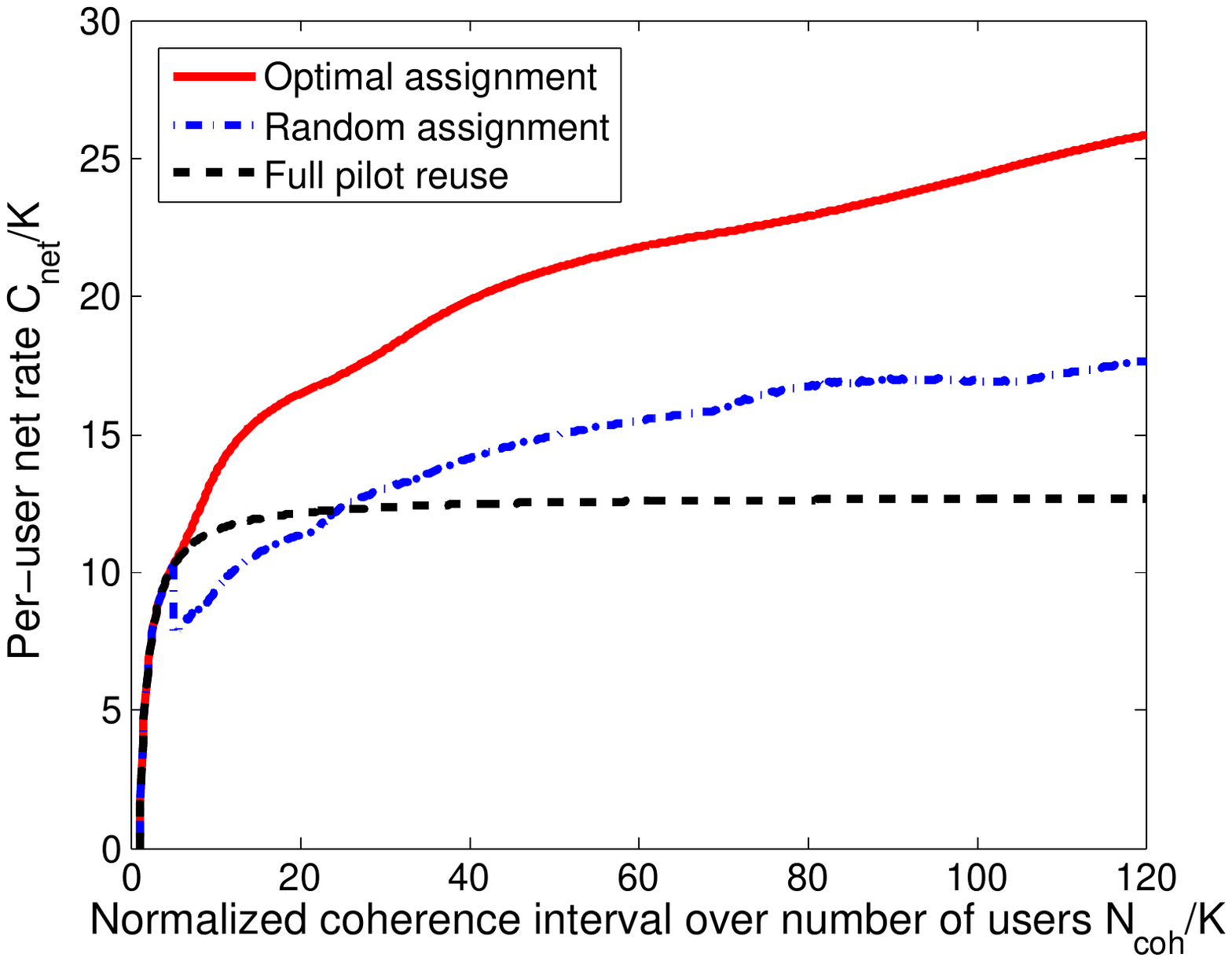}
    \caption{Per-user net rate versus $N_{coh}/K$}
    \label{Fig:K=14 Simulation}
    
\end{figure}

Fig. \ref{Fig:K=14 Simulation} shows the plots of the per-user net rates $C_{net}/K$
versus 
$N_{coh}/K$. These results are actually found for $K=14$, but very similar numerical results 
are obtained for the optimal scheme irrespective of the particular values of $K$ while
the results for the full pilot reuse are identical across all values of $K$. As a case in point,
it can be seen that the plots in Fig. \ref{Fig:K=1 Simulation} obtained for 
$K=1$ are nearly identical to those in Fig. \ref{Fig:K=14 Simulation} corresponding to $K=14$.

In fact, it can be shown that for
all valid pilot vectors with only one non-zero component, $\mathbf{p}=(K, 0, \cdots, 0)$, $(0, 3K, 0, \cdots, 0)$, $(0, 0, 9K, 0, \cdots, 0)$, $\cdots$, $(0, \cdots, 0, \frac{LK}{3})$,
the plot of $C_{net}/K$ verses $N_{coh}/K$ does not change with $K$. 
For the remaining pilot assignment vectors, the plot changes very little 
across different values of $K$. As can be seen, using the optimal pilot assignment scheme 
the per user net rate improves 
substantially with increasing $N_{coh}/K$, relative to the full reuse scheme.

\begin{figure}[!t]
\centering
        \includegraphics[height=60mm]{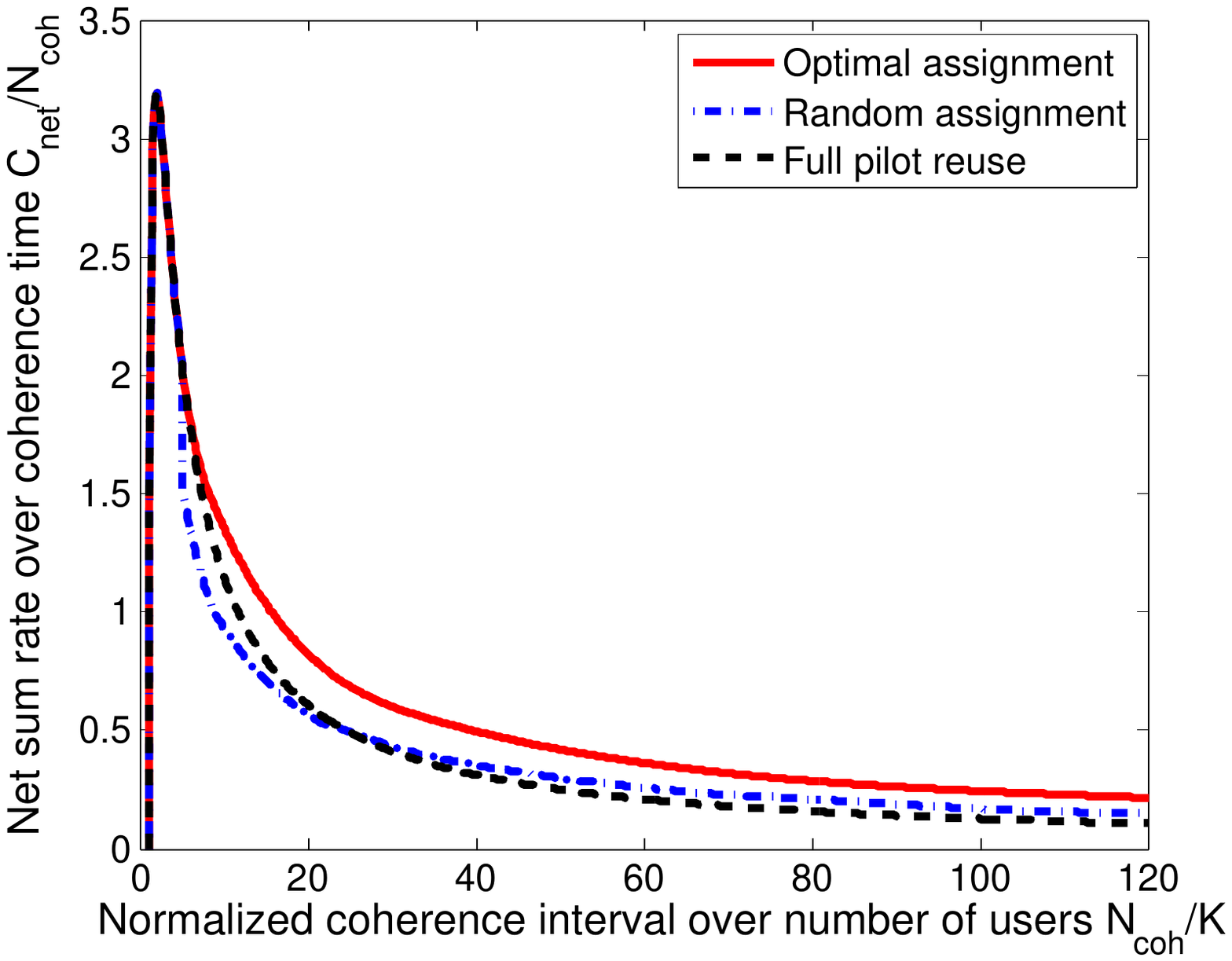}
    \caption{$C_{net}/N_{coh}$ versus $N_{coh}/K$ for different pilot assignment schemes}
    \label{Fig:K=14 Simulation2}
    
\end{figure}

Fig. \ref{Fig:K=14 Simulation2} shows $C_{net}/N_{coh}$ versus $N_{coh}/K$ for 
optimal, full reuse and random pilot assignment schemes. 
Again, $K=14$ is used to generate these plots ($L=81$ throughout this paper), 
but the plots do not change noticeably 
for different values of $K$. 
These plots give an insight 
into how the net sum rate changes as $K$ decreases while $N_{coh}$ is held fixed.
Notice that the maximum net sum-rate occurs at $N_{coh}/K=2$ 
and at this point the optimal scheme reduces to full reuse,
consistent with the
Marzetta's analysis \cite{ref:Marzetta2}.
However, the message here is, again, that when we do not have a control over 
$N_{coh}/K=2$, the optimal pilot assignment strategy may have a substantial net sum-rate advantage 
over full pilot reuse.

To appreciate how large $N_{coh}/K$ can be in some real-world scenarios,
take an indoor office wireless channel with $T_{coh}=50$ microsec and $T_{del}=50$ nsec, yielding 
$N_{coh}=1000$. If the user density cannot be allowed to be more than $K=20$ users per cell, 
we would be focusing on $N_{coh}/K=50$ and at this point, the optimal scheme gives  a 67\% net sum-rate improvement over full pilot reuse. As another example, consider an urban outdoor environment with a fairly high user mobility giving rise to a 1 msec coherence time interval. With
a 2 microsec delay spread, for example, this gives $N_{coh}=500$, and 
assuming not more than $K=25$ users are to be served, we are interested in the net sum-rates at $N_{coh}/K=20$. From either Fig. \ref{Fig:K=14 Simulation} or Fig. \ref{Fig:K=14 Simulation2}, we see that a net throughput improvement of $35\%$ is possible via optimal pilot assignment, relative to full pilot reuse. 

\section{Conclusion}
In massive MIMO systems with interfering cells, when the coherence time interval $N_{coh}$
and the number of users $K$ are given, finding the appropriate portion of the coherence interval to
be allocated to channel training is not trivial.
This paper has provided an analytical solution to finding the optimal training time 
for any given ratio of $N_{coh}/K$
and along the way showed that the net throughput could be improved in general by
allowing neighboring cells to use different sets of pilot sequences. 
Assuming hexagonal cells and equi-distance hierarchical partitioning, an optimal 
pilot assignment strategy has been identified that gives substantial throughput advantages 
relative to random pilot assignment or full pilot reuse 
when the ratio $N_{coh}/K$ is sufficiently large.
Finally, we add that it would be interesting to further explore 
pilot assignment strategies when the objective is not about
maximizing the sum rate but rather on
guaranteeing some minimal performance level to all users 
or maximizing a weighted sum rate to prioritize the services.  

\end{document}